\documentclass[sigconf, nonacm]{acmart}

\usepackage{subcaption}
\usepackage{hyperref}
\usepackage{cite}
\usepackage{amsmath,amsfonts}
\usepackage{algorithmic}
\usepackage{graphicx}
\usepackage{textcomp}
\usepackage{xcolor}
\usepackage{multirow}
\usepackage{tikz}
\usetikzlibrary{fit}
\usetikzlibrary{positioning,arrows.meta,shapes.geometric,backgrounds}

\AtBeginDocument{%
  \providecommand\BibTeX{{%
    \normalfont B\kern-0.5em{\scshape i\kern-0.25em b}\kern-0.8em\TeX}}}

\copyrightyear{2026}
\acmYear{2026}
\setcopyright{cc}
\setcctype{by}
\acmConference[WSESE '26]{3rd International Workshop on Methodological Issues with Empirical Studies in Software Engineering }{April 12--18, 2026}{Rio de Janeiro, Brazil}
\acmBooktitle{3rd International Workshop on Methodological Issues with Empirical Studies in Software Engineering (WSESE '26), April 12--18, 2026, Rio de Janeiro, Brazil}
\acmPrice{}
\acmDOI{10.1145/3786149.3788308}
\acmISBN{979-8-4007-2382-7/2026/04}

\begin{document}

\acmPrice{15.00}
\acmDOI{10.1145/3530019.3535306}
\acmISBN{978-1-4503-9613-4/22/06}

\makeatletter
\newcommand*\bigcdot{\mathpalette\bigcdot@{.5}}
\newcommand*\bigcdot@[2]{\mathbin{\vcenter{\hbox{\scalebox{#2}{$\m@th#1\bullet$}}}}}
\makeatother

\title{Towards an OSF-based Registered Report Template for Software Engineering Controlled Experiments} 

\author{Ana B. M. Bett}
\affiliation{
  \institution{State University of Maringá (UEM)}
  \city{Maringá}
  \country{Brazil}}
\email{anabmbett@gmail.com}

\author{Thais S. Nepomuceno}
\affiliation{
  \institution{State University of Maringá (UEM)}
  \city{Maringá}
  \country{Brazil}}
\email{thais.nepomuceno1@gmail.com}

\author{Edson OliveiraJr}
\affiliation{
  \institution{State University of Maringá (UEM)}
  \city{Maringá}
  \country{Brazil}}
\email{edson@din.uem.br}

\author{Maria Teresa Baldassare}
\affiliation{
  \institution{University of Bari}
  \city{Bari}
  \country{Italy}}
\email{mariateresa.baldassarre@uniba.it}

\author{Valdemar V. Graciano Neto}
\affiliation{
  \institution{Federal University of Goiás (UFG)}
  \city{Goiânia}
  \country{Brazil}}
\email{valdemarneto@ufg.br}

\author{Marcos Kalinowski}
\affiliation{
  \institution{PUC-Rio}
  \city{Rio de Janeiro}
  \country{Brazil}}
\email{kalinowski@inf.puc-rio.br}

\renewcommand{\shortauthors}{Bett et al.}

\begin{abstract}
\textit{Context:} The empirical software engineering (ESE) community has contributed to improving experimentation over the years. However, there is still a lack of rigor in describing controlled experiments, hindering reproducibility and transparency. Registered Reports (RR) have been discussed in the ESE community to address these issues. A RR registers a study’s hypotheses, methods, and/or analyses before execution, involving peer review and potential acceptance before data collection. This helps mitigate problematic practices such as p-hacking, publication bias, and inappropriate \textit{post hoc} analysis.
\textit{Objective:} This paper presents initial results toward establishing an RR template for Software Engineering controlled experiments using the Open Science Framework (OSF).
\textit{Method:} We analyzed templates of selected OSF RR type in light of documentation guidelines for controlled experiments.
\textit{Results:} The observed lack of rigor motivated our investigation of OSF-based RR types. Our analysis showed that, although one of the RR types aligned with many of the documentation suggestions contained in the guidelines, none of them covered the guidelines comprehensively. The study also highlights limitations in OSF RR template customization.
\textit{Conclusion:} Despite progress in ESE, planning and documenting experiments still lack rigor, compromising reproducibility. Adopting OSF-based RRs is proposed. However, no currently available RR type fully satisfies the guidelines. Establishing RR-specific guidelines for SE is deemed essential.
\end{abstract}

\maketitle

\section{Introduction}

Software engineering (SE) research still faces numerous challenges, one of which consists of obtaining and managing research artifacts \citep{OliveiraJr_et_al2024}. Rigorous research is mainly based on transparency, credibility, and reproducibility, which are pillars in constructing and evaluating robust theories for the area's evolution \citep{OliveiraJr_et_al2021}. Open Science principles and practices can provide a solid foundation for achieving these goals\footnote{\url{https://unesdoc.unesco.org/ark:/48223/pf0000379949}}~\citep{mendez2020open}.

In recent years, several major SE conferences, such as ICSE\footnote{\url{https://conf.researchr.org/track/icse-2025/icse-2025-artifact-evaluation}} and FSE\footnote{\url{https://conf.researchr.org/track/fse-2025/fse-2025-artifacts}}, have introduced Artifact Tracks. These tracks focus on evaluating and promoting the quality and availability of research artifacts, such as software tools, data, and frameworks that support the reproducibility and validity of research findings. The conferences share common goals and inclusion criteria: artifacts must be functional, reusable, and publicly available in persistent repositories with proper documentation. By incorporating these tracks, the conferences highlight the importance of artifacts in ensuring the transparency, reliability, and reproducibility of empirical studies in software engineering.

To emphasize artifact quality, many conferences, including ICSE and FSE, collaborate with the ACM’s badging system. This system awards badges to artifacts that meet specific criteria, such as functionality and reusability. These badges recognize high-quality artifacts that particularly enhance the reproducibility and reusability of research. Additionally, the ACM badges encourage the community to prioritize transparency and the open sharing of research results, ensuring that artifacts are accessible and valuable for researchers seeking to replicate or build upon existing findings.

Although reproducibility and reliability are crucial, many Software Engineering studies still face challenges, including issues with replicability, biases in results, and incomplete research artifacts. 

In response to these challenges, Registered Reports (RRs) have emerged as a promising solution, as they enhance research quality by requiring hypotheses and study protocols to be established prior to data collection, thereby reducing bias and increasing transparency. RRs are peer-reviewed, which straightforwardly improves such characteristics. RRs are being adopted at several SE conferences, including ICSE, ESEM, ICSME, and MSR.

Recent studies have addressed this topic, including that of \citet{ernst2023registered}. However, this topic remains crucial for the reproducibility, reliability, and transparency of empirical studies.

In this paper, we focus on the following research question: ``\textbf{how can we design an RR template for SE controlled experiments based on existing guidelines?}''. To do so, we propose and discuss an initial template based on the Open Science Framework (OSF) platform. OSF offers various RR templates for different kinds of empirical studies. Therefore, we analyzed and compared all of them to determine how well they align with the specific guidelines for SE-controlled experiments compiled by \citet{Jedlitschka_et_al2008}. We also discuss the limitations of this study, share the lessons learned, and suggest potential future actions.

This paper is organized as follows: Section \ref{sec:guidelines} presents the analyzed controlled experimentation guidelines; Section \ref{sec:rr} presents essential background on Registered Reports and the Open Science Framework platform used for proposing an RR template; Section \ref{sec:results} presents our methodology, the proposed initial RR template, and discusses lessons learned and limitations; Section \ref{sec:actions} presents prospective actions based on the proposed template; and Section \ref{sec:conclusion} presents final remarks.

\section{Guidelines for Documenting Controlled Experiments and RR in SE}
\label{sec:guidelines}

\citet{Jedlitschka_et_al2008} present a set of guidelines for experiments in SE. These guidelines provide detailed guidance on the expected content of sections and subsections for reporting controlled experiments and \textit{quasi}-experiments.
The guidelines are organized into sections, each one focusing on a core component of controlled experiment reporting in SE. These sections cover structured abstract, introduction, experiment planning, execution, analysis, interpretation, and conclusions. Table \ref{tab:guidelines} summarizes these guidelines by section and scope.

\begin{table}[!h]
    \centering
    \caption{Experiment Guidelines by \citet{Jedlitschka_et_al2008}}
    \label{tab:guidelines}
    \resizebox{.48\textwidth}{!}{
    \begin{tabular}{c|l|l|p{6cm}}
    \hline \hline
    \textbf{ID} & \textbf{Section} & \textbf{Sub-Section} & \textbf{Scope} \\
    \hline \hline
    G1  & Title                    &                      & \textless title\textgreater + "- A controlled experiment" \\
    G2  & Authorship               &                      & Does it include contact information? \\
    G3  & Structured Abstract       & Background           & Why is this research important? \\
    G4  & Structured Abstract       & Objective            & What is the question addressed with this research? \\
    G5  & Structured Abstract       & Methods              & What is the statistical context and methods applied? \\
    G6  & Structured Abstract       & Results              & What are the main findings? Practical implications? \\
    G7  & Structured Abstract       & Limitations          & What are the weaknesses of this research? \\
    G8  & Structured Abstract       & Conclusions          & What is the conclusion? \\
    G9  & Keywords                 &                      & Areas of research, treatments, dependent variables, and study type \\
    G10 & Introduction             & Problem Statement    & What is the problem? Where does it occur? Who has observed it? Why is it important? \\
    G11 & Introduction             & Research Objectives  & Analyze \textless Object(s) of study\textgreater for the purpose of \textless purpose\textgreater \\
    G12 & Introduction             & Context              & What are the environmental factors that impact generalizability? \\
    G13 & Related Work             &                      & How does it relate to existing research or state of the practice? \\
    G14 & Experiment Planning      & Goals                & Formalization of goals, defining important constructs \\
    G15 & Experiment Planning      & Experimental Units   & From which population will the sample be drawn? \\
    G16 & Experiment Planning      & Experimental Material & Which objects are selected and why? \\
    G17 & Experiment Planning      & Tasks                & Which tasks have to be performed by the subjects? \\
    G18 & Experiment Planning      & Hypotheses, Variables & Formalization of hypotheses and variables \\
    G19 & Experiment Planning      & Design               & What type of experimental design was chosen? \\
    G20 & Experiment Planning      & Procedure            & How will the experiment (data collection) be performed? \\
    G21 & Experiment Planning      & Analysis Procedure   & How will the data be analyzed? \\
    G22 & Execution                &                      & Main purpose: describe any deviations from the plan \\
    G23 & Execution                & Preparation          & What was done to prepare for the execution of the experiment? \\
    G24 & Execution                & Deviations           & Describe any deviations from the plan, e.g., data collection \\
    G25 & Analysis                 & Descriptive Statistics & What are the results from descriptive statistics? \\
    G26 & Analysis                 & Data Set Reduction   & Was it necessary to reduce the dataset? Why and how? \\
    G27 & Analysis                 & Hypothesis Testing   & How was the data evaluated? Was the model validated? \\
    G28 & Interpretation           & Evaluation of Results & Explain the results in relation to earlier research \\
    G29 & Interpretation           & Threats to Validity   & How is the validity of the results assured? Discuss potential biases \\
    G30 & Interpretation           & Inferences           & Inferences drawn from the data to more general conditions \\
    G31 & Interpretation           & Lessons Learned      & What was learned during the experiment? \\
    G32 & Conclusions & Summary    & Provide a concise summary of the research and results \\
    G33 & Conclusions & Impact     & Description of impacts on cost, schedule, and quality \\
    G34 & Conclusions & Future Work & What other experiments could be run to further investigate the findings? \\
    G35 & Acknowledgements         &                      & Acknowledge sponsors, participants, and contributors \\
    G36 & References               &                      & All cited literature must be presented in the requested format \\
    G37 & Appendices               &                      & Material, raw data, detailed analyses for reproducibility \\
    \hline \hline
    \end{tabular}
    }
\end{table}

The first element emphasizes the importance of a clear, informative title (\textbf{G1}) that explicitly states that the work is a controlled experiment, as well as complete authorship details (\textbf{G2}), including contact information for accountability and follow-up. The structured abstract (\textbf{G3--G8}) plays a central role, as it must present the background and motivation of the study (\textbf{G3}), define its objectives and research questions (\textbf{G4}), describe the statistical context and methods employed (\textbf{G5}), summarize the main results and their practical implications (\textbf{G6}), acknowledge the study’s limitations (\textbf{G7}), and provide a clear conclusion (\textbf{G8}). This standardized format makes it easier for readers and reviewers to understand, compare, and reuse the reported evidence.

Keywords (\textbf{G9}) must be carefully chosen to reflect the domain, treatments, dependent variables, and study type, supporting efficient indexing in scientific databases. In the introduction (\textbf{G10--G12}), researchers should present a well-defined problem statement (\textbf{G10}), explain the research objectives (\textbf{G11}), and clarify the context in which the study arises, including environmental and organizational factors that may affect generalizability (\textbf{G12}). The introduction should also be complemented by a discussion of related work (\textbf{G13}), which situates the experiment within prior studies and identifies its novelty or added contribution.

The experiment planning section (\textbf{G14--G21}) provides the foundation for methodological rigor. Goals must be formalized (\textbf{G14}), while the description of experimental units (\textbf{G15}), materials (\textbf{G16}), and tasks (\textbf{G17}) clarifies the scope of the study. Hypotheses and variables should be explicitly defined (\textbf{G18}), and the chosen experimental design explained (\textbf{G19}). Furthermore, the procedures for execution and data collection (\textbf{G20}) and the analysis procedure (\textbf{G21}) should be described in sufficient detail to enable replication and critical evaluation.

Execution (\textbf{G22--G24}) is then reported with emphasis on preparation activities (\textbf{G23}) and transparent documentation of any deviations from the original plan (\textbf{G24}). This is followed by the analysis section (\textbf{G25--G27}), where descriptive statistics (\textbf{G25}) provide an overview of the data (dataset reduction is justified if applied (\textbf{G26}), and hypothesis testing (\textbf{G27}), which is performed for validating or rejecting the proposed assumptions. The interpretation of results (\textbf{G28--G31}) links findings to earlier research (\textbf{G28}), discusses threats to validity and potential biases (\textbf{G29}), and formulates inferences that extend to broader contexts (\textbf{G30}). Lessons learned (\textbf{G31}) are also emphasized, capturing insights gained during the study.

Finally, the conclusions (\textbf{G32--G34}) must concisely summarize the research and results (\textbf{G32}), highlight their impact on aspects such as cost, schedule, and quality (\textbf{G33}), and suggest directions for future work (\textbf{G34}). The closing elements of acknowledgments (\textbf{G35}), references (\textbf{G36}), and appendices (\textbf{G37}) strengthen the credibility and reproducibility of the study by recognizing contributions, providing properly formatted citations, and sharing supplementary material, raw data, and detailed analyses.

\section{Registered Reports (RR) in Software Engineering and the OSF Framework}
\label{sec:rr}

This section presents essential concepts on RR and the OSF framework.

\subsection{Registered Reports}

Science must be transparent and accessible, regardless of study results. Despite its theoretical and empirical value, researchers often lack the motivation to conduct replications or report negative results \citep{nosek2014registered, crawford1990peer}. Access to the work of other researchers is necessary for evaluating, replicating, and advancing knowledge~\citep{crawford1990peer}.

The traditional scientific process involves sharing research methods and findings mainly through journal articles, theses, and dissertations. However, many software engineering research artifacts are not shared in a way that allows them to be reproduced \citep{ernst2023registered}. To address this issue, it may be helpful to preregister a study protocol, a practice successfully employed in fields such as medicine for decades. Preregistration, also known as a Registered Report (RR), involves openly sharing the study protocol and discussing its validity \citep{silva2017}.

RRs are scientific publications that precede the publication process. They include a detailed research protocol, research questions, a literature review, and specifications for the data analysis procedures to be used. This practice has recently become present in some computer science conferences and journals \citep{ernst2023registered}.

RRs have two stages: Stage 1, where the protocol is registered and formalized, and Stage 2, which reports the execution of the study in accordance with the protocol. Any deviations or changes must be reported (in general, we expect deviations to be minimal).

An RR is a valuable tool to ensure a certain level of quality in study design. For example, it ensures that the hypotheses of a confirmatory study are predefined rather than being defined after analyzing the data to fit the results. Researchers define their research questions, the reasons for pursuing the research, and exactly how they will try to answer their questions \citep{mendez2020open}.

An RR is based on the philosophy that it is essential to focus on research planning and to evaluate the results without bias, to minimize distortions in scientific findings \citep{chambers2022past}. This method involves initiating the evaluation before any data is collected, enabling early feedback that can save time and effort. Seeking opinions from individuals outside of the research group can avoid potential issues with research practices and provide valuable early feedback to researchers \citep{ernst2023registered}.

The process of preparing an RR involves two phases \citep{chambers}. In the first phase, authors submit their research questions, theory, detailed methods, analysis plans, and any preliminary data as required. After a thorough analysis and review based on specific criteria, proposals that are favorably evaluated receive a conditional acceptance, which means that the venue (journal or conference) commits to publishing the final article regardless of whether the hypotheses are supported, as long as the authors adhere to their approved protocols and interpret the results based on the evidence. The approved protocol, results, and discussion are included in the second phase.

The advantages of RRs include shareable protocols for research replication, a focus on research rather than publication, and greater rigor in reporting. However, this practice may also have disadvantages, such as requiring greater effort from researchers, currently facing limited acceptance by journals, and not applying to all research strategies \citep{ernst2023registered}. Multiple RRs for various areas are accessible on the PCI Registered Reports Community\footnote{\url{https://rr.peercommunityin.org/}} website.
An example of an RR for a SE controlled experiment can be found at \url{https://rr.peercommunityin.org/articles/rec?id=746}.

\subsection{The Open Science Framework (OSF)}
\label{sec:osf}

The Open Science Framework\footnote{\url{https://osf.io/}} (OSF) platform is designed to assist researchers, whether they are working on collaborative projects or not. This platform connects data, pre-publications, and research project progress, offering opportunities for publishing, categorizing, and obtaining community feedback. It has the following main features:

\begin{itemize}
    \item The functionality dedicated to \textbf{Institutions} is designed for universities and research institutes, aiming to foster a community among researchers. It simplifies the management of research projects by providing features to organize, document, and securely share research materials and data;
    \item The \textbf{Meetings} feature streamlines the process of sharing posters and presentations openly at academic conferences. Conference organizers can create a dedicated page on OSF for participants to submit their work;
    \item \textbf{Preprints} on the OSF platform are an effective and agile way for researchers to share their preliminary results with the academic community before submitting them for peer review in scientific journals; and
    \item  The \textbf{Registries} feature provides a centralized and accessible platform for academic community members to share and explore studies conducted by others. It enables the investigation of overviews, metadata, files, resources, components, links, analytics, and comments.
\end{itemize}

An exemplary general OSF-based RR can be found at \url{https://doi.org/10.17605/OSF.IO/JSZNK}. The resulting publication of this RR is available at \url{https://doi.org/10.31234/osf.io/cdgyh}.

\section{An OSF-based RR Template Proposal}
\label{sec:results}

In this section, we demonstrate how we developed an RR Stage-1 template for Software Engineering controlled experiments based on the OSF RR types and documentation items.

We started by analyzing and comparing each RR type with each other (Section \ref{sec:osf_rrs}). We familiarized ourselves with the RR types available in the OSF platform and their purpose. Subsequently, we excluded RR types that were unsuitable for controlled experiments. In the final stage (Section \ref{sec:mapping}), we compared the recommendations \citet{Jedlitschka_et_al2008} with each remaining OSF RR type. We will analyze and discuss the lessons learned and the limitations of the template proposal in Section \ref{sec:lessons}.

\subsection{The OSF Registered Report Types}
\label{sec:osf_rrs}

The OSF platform offers 11 types of RRs to formalize and enhance transparency in scientific research, as shown below. Every kind of RR possesses distinct characteristics that cater to various research needs.

The \textbf{OSF Preregistration} (RR.1) type is suitable for providing a detailed, transparent narrative of a study from initial planning to final results.

The \textbf{Open-Ended Registration} (RR.2) consists of a summary narrative of the study information, allowing the inclusion of supplementary files or additional information. 

\textbf{Qualitative Preregistration} (RR.3) is intended to guide qualitative research studies by providing detailed information about objectives, design, data collection, and analysis.

\textbf{Secondary Data Preregistration} (RR.4) is crucial for conducting analyses based on existing data, as it provides information on research questions, data description, variables, and planned statistical analyses.

The \textbf{Generalized Systematic Review Registration} (RR.5) is intended for preregistration of systematic reviews, providing information on methods, search strategies, screening, data extraction, and synthesis.

The \textbf{Registered Report Protocol Preregistration} (RR.6) Pre-registration type is required for studies that have received in-principle acceptance from journals offering Registered Reports. 

The \textbf{OSF-Standard Pre-Data Collection Registration} (RR.7) is crucial for providing information on the status of data collection and data access. 

\textbf{Preregistration Template from AsPredicted.org}  (RR.8) is intended to ensure a complete description of the study, covering aspects such as questions/hypotheses, variables, statistical analyses, and other relevant details. 

The \textbf{Replication Recipe (Brandt et al., 2013): Post-Completion} (RR.9) provides a framework for detailed recording of planned methodological procedures, ensuring transparency and reproducibility in replication. 

The \textbf{Replication Recipe (Brandt et al., 2013): Pre-Registration} (RR.10) is designed to register replication attempts of pre-existing studies, providing details about the objectives, methods, and planned analyses.

\textbf{Preregistration in Social Psychology} (RR.11) is explicitly intended for the pre-registration of studies in Social Psychology, covering aspects such as hypotheses, variables, methods, analyses, and other relevant details. 

Table \ref{tab:relacao_rr_itens} summarizes the documentation items for RR.1, RR.3, RR.10, and RR.11 OSF Registered Report Types. The complete table, which includes all 11 RR types, is available at \url{https://doi.org/10.5281/zenodo.11114264}. We excluded the types RR.2, RR.4, RR.5, RR.6, RR.7, RR.8, and RR.9 because they are unsuitable for controlled experiments.

\begin{table}[!h]
    \centering
    \caption{Documentation Items of RR.1, RR.3, RR.10, and RR.11 OSF Registered Report Types}
    \label{tab:relacao_rr_itens}

    \resizebox{0.48\textwidth}{!}{
    \begin{tabular}{l|l|c|c|c|c}
    \hline \hline 

        \textbf{Category} &	\textbf{Item}	&	\textbf{RR.1}	&	\textbf{RR.3}	&	\textbf{RR.10}	&	\textbf{RR.11} \\
         \hline  \hline 

        Metadata	&	Title	&	\checkmark	&	\checkmark	&	\checkmark	&	\checkmark	\\	\hline
        Metadata	&	Description	&	\checkmark	&	\checkmark	&	\checkmark	&	\checkmark	\\	\hline
        Metadata	&	Contributors	&	\checkmark	&	\checkmark	&	\checkmark	&	\checkmark	\\	\hline
        Metadata	&	Category	&	\checkmark	&	\checkmark	&	\checkmark	&	\checkmark	\\	\hline
        Metadata	&	Licensing	&	\checkmark	&	\checkmark	&	\checkmark	&	\checkmark	\\	\hline
        Metadata	&	Subjects	&	\checkmark	&	\checkmark	&	\checkmark	&	\checkmark	\\	\hline
        Metadata	&	Tags	&	\checkmark	&	\checkmark	&	\checkmark	&	\checkmark	\\	\hline
        Study Information	&	Specific, Concise, and Testable Hypotheses	&	\checkmark	&		&		&		\\	\hline
        Study Information	&	Research Aims	&		&	\checkmark	&		&		\\	\hline
        Study Information	&	Type of Aim 	&		&	\checkmark	&		&		\\	\hline
        Study Information	&	Research question(s)	&		&	\checkmark	&		&		\\	\hline
        Study Information	&	Anticipated Duration	&		&	\checkmark	&		&		\\	\hline
        Design Plan	&	Study Type	&	\checkmark	&		&		&		\\	\hline
        Design Plan	&	Blinding Type	&	\checkmark	&		&		&		\\	\hline
        Design Plan	&	Factors and Treatments	&	\checkmark	&		&		&		\\	\hline
        Design Plan	&	Randomization	&	\checkmark	&		&		&		\\	\hline
        Design Plan	&	Study Design	&		&	\checkmark	&		&		\\	\hline
        Design Plan	&	Sampling and Case Selection Strategy	&		&	\checkmark	&		&		\\	\hline
        Data Collection	&	Data Source(s) and Data Type(s)	&		&	\checkmark	&		&		\\	\hline
        Data Collection	&	Data Collection Methods	&		&	\checkmark	&		&		\\	\hline
        Data Collection	&	Data Collection Tools, Instruments or Plans	&		&	\checkmark	&		&		\\	\hline
        Data Collection	&	Stopping Criteria	&		&	\checkmark	&		&		\\	\hline
        Sampling Plan	&	Existing Data Use	&	\checkmark	&		&		&		\\	\hline
        Sampling Plan	&	Explanation of Existing Data	&	\checkmark	&		&		&		\\	\hline
        Sampling Plan	&	Data Collection Procedures	&	\checkmark	&		&		&		\\	\hline
        Sampling Plan	&	Sample Size and Rationale	&	\checkmark	&		&		&		\\	\hline
        Sampling Plan	&	Stopping Rule	&	\checkmark	&		&		&		\\	\hline
        Variables	&	Manipulated Variables	&	\checkmark	&		&		&		\\	\hline
        Variables	&	Measurable Variables	&	\checkmark	&		&		&		\\	\hline
        Variables	&	Combination of Variables in Index	&	\checkmark	&		&		&		\\	\hline
        Analysis Plan	&	Statistical Models	&	\checkmark	&		&		&		\\	\hline
        Analysis Plan	&	Data Transformation, Centering, and Recoding	&	\checkmark	&		&		&	\checkmark	\\	\hline
        Analysis Plan	&	Coding Scheme for Categorical Variables	&	\checkmark	&		&		&		\\	\hline
        Analysis Plan	&	Inference Criteria	&	\checkmark	&		&		&		\\	\hline
        Analysis Plan	&	Data Exclusion	&	\checkmark	&		&		&	\checkmark	\\	\hline
        Analysis Plan	&	Missing Data	&	\checkmark	&		&		&		\\	\hline
        Analysis Plan	&	Exploratory Analysis	&	\checkmark	&		&		&		\\	\hline
        Analysis Plan	&	Approach	&		&	\checkmark	&		&		\\	\hline
        Analysis Plan	&	Process	&		&	\checkmark	&		&		\\	\hline
        Analysis Plan	&	Credibility Strategies	&		&	\checkmark	&		&		\\	\hline
        Analysis Plan	&	Reliability and Robustness Testing	&		&		&		&	\checkmark	\\	\hline
        Analysis Plan	&	Revariables	&		&		&		&	\checkmark	\\	\hline
        Analysis Plan	&	Statistical Technique	&		&		&		&	\checkmark	\\	\hline
        Analysis Plan	&	Variable Roles	&		&		&		&	\checkmark	\\	\hline
        Analysis Plan	&	Covariate Rationale	&		&		&		&	\checkmark	\\	\hline
        Analysis Plan	&	Other Techniques	&		&		&		&	\checkmark	\\	\hline
        Analysis Plan	&	Method of Correction	&		&		&		&	\checkmark	\\	\hline
        Analysis Plan	&	Assumptions of Analysis	&		&		&		&	\checkmark	\\	\hline
        Analysis Plan	&	Data Collection	&		&		&		&	\checkmark	\\	\hline
        Analysis Plan	&	Data Observation	&		&		&		&	\checkmark	\\	\hline
        Analysis Plan	&	Start and End Dates	&		&		&		&	\checkmark	\\	\hline
        Analysis Plan	&	Additional Comments	&		&		&		&	\checkmark	\\	\hline
        The Nature of the Effect	&	Description	&		&		&	\checkmark	&		\\	\hline
        The Nature of the Effect	&	Replication Importance	&		&		&	\checkmark	&		\\	\hline
        The Nature of the Effect	&	Effect Size	&		&		&	\checkmark	&		\\	\hline
        The Nature of the Effect	&	Confidence Interval	&		&		&	\checkmark	&		\\	\hline
        The Nature of the Effect	&	Sample Size	&		&		&	\checkmark	&		\\	\hline
        The Nature of the Effect	&	Original Study Conducted	&		&		&	\checkmark	&		\\	\hline
        The Nature of the Effect	&	Replication Confidence	&		&		&	\checkmark	&		\\	\hline
    \end{tabular}
    }
\end{table}

The ``Metadata'' category is common to all 11 RR types. Certain identical items are under different categories, such as ``Exclusion Criteria'' under ``Analysis and Replication  Evaluation'' for RR.10 and ``Methods'' for RR.11.

RR.1 and RR.3 are the types of RR that have the most additional items (Category) compared to the other types. The main distinction between these two types is that RR.3 is used for qualitative research studies, while RR.1 concerns a more general study.

\subsection{Mapping of Experimentation Guidelines and OSF RRs}
\label{sec:mapping}

Table \ref{tab:relacao_rr_diretrizes} shows how the experiment guidelines map to the type of OSF RR, considering which guidelines are satisfied for each RR type.

\begin{table}[!h]
    \centering
    \caption{Types of OSF Registered Reports and Respective Controlled Experimentation Guidelines}
    \label{tab:relacao_rr_diretrizes}   
    \resizebox{0.49\textwidth}{!}{
        \begin{tabular}{c|p{2cm}|p{2cm}|p{2cm}|p{2cm}|c}

        \hline \hline
        \multirow{2}{*}{\textbf{OSF RR Type}} & \multicolumn{5}{c}{\textbf{Controlled Experimentation Guidelines}}\\ 
        
        \cline{2-6} & \textbf{Doc.} & \textbf{Planning} & \textbf{Operation} & \textbf{Analysis} & \textbf{Count}\\
        \hline   \hline     
        
        RR.1  & G.1, G.2, G.3, G.4, G.5, G.6, G.10, G.11, G.12, G.13, G.36, G.37 & G.14, G.15, G.16, G.17, G.18, G.19, G.20, G.21 & G.23, G.24 & G.25, G.26, G.27, G.28, G.29, G.30, G.32, G.33, G.34 & 31 \\\hline 
        RR.2  & G.1, G.2, G.13 & G.15 & G.23, G.24 &  & 6 \\\hline 
        RR.3  & G.1, G.2, G.3, G.4, G.5, G.6, G.7, G.8, G.9, G.10, G.11, G.12 & G.14, G.15, G.16, G.17, G.18, G.19, G.20, G.21 & G.22, G.23, G.24 & G.25, G.26, G.27, G.28, G.29, G.30, G.31, G.32, G.33, G.34 & 33 \\\hline 
        RR.4  & G.1, G.2, G.4, G.5, G.11, G.13 & G.14, G.15, G.16, G.17, G.18, G.19, G.20, G.21 & G.22, G.23 & G.25, G.26, G.27, G.28, G.29 & 21 \\\hline 
        RR.5  & G.1, G.2, G.3, G.10, G.13, G.37 &  &  &  & 6 \\\hline 
        RR.6  & G.1, G.2, G.35, G.36, G.37 &  & G.23, G.24 &  & 7 \\\hline 
        RR.7  & G.1, G.2, G.36, G.37 &  & G.24 &  & 5 \\\hline 
        RR.8  & G.1, G.2, G.3, G.4, G.5, G.7, G.10, G.37 & G.14, G.15, G.18, G.19, G.20, G.21 & G.22, G.23, G.24 & G.26, G.27, G.29, G.30 & 21 \\\hline     
        RR.9  & G.1, G.2, G.4, G.5, G.6, G.7, G.8, G.37 & G.14, G.15, G.16, G.18, G.20, G.21 &  &  & 15 \\\hline 
        RR.10 & G.1, G.2, G.3, G.4, G.5, G.6, G.7, G.8, G.10, G.11, G.12, G.37 & G.14, G.15, G.16, G.17, G.18, G.19, G.20, G.21 & G.23, G.24 & G.27, G.28, G.29 & 25 \\\hline 
        RR.11 & G.1, G.2, G.3, G.4, G.5, G.6, G.8, G.10, G.11, G.12, G.37 & G.14, G.15, G.16, G.17, G.18, G.19, G.20, G.21 & G.23, G.24 & G.25, G.26, G.27, G.28, G.29, G.30, G.32, G.34 & 29 \\\hline 
        \hline
    \end{tabular}
    }
\end{table}

Table \ref{tab:relacao_diretrizes_rr} presents a different perspective of Table \ref{tab:relacao_rr_diretrizes} in terms of summarizing the comparison of RR types with respect to the guidelines.

\begin{table*}[!h]
    \centering
    \tiny
    \caption{Experimentation Guidelines per OSF Registered Report Types}
    \label{tab:relacao_diretrizes_rr}
    \resizebox{0.99\textwidth}{!}{ 
        \begin{tabular}{l|c|c|c|c|c|c|c|c|c|c|c|c|c}

        \hline \hline
         \multicolumn{2}{c|}{\textbf{Experiment Guidelines}} & \multicolumn{11}{c|}{\textbf{Types of OSF Registered Reports}} & \multirow{2}{*}{\textbf{Count}} \\ 
        
        \cline{1-2} \cline{3-13} 
        
        \textbf{Phase} & \textbf{ID} & \textbf{RR.1} & \textbf{RR.2} & \textbf{RR.3} & \textbf{RR.4} & \textbf{RR.5} & \textbf{RR.6} & \textbf{RR.7} & \textbf{RR.8} & \textbf{RR.9} & \textbf{RR.10} & \textbf{RR.11} & ~\\
         \hline \hline 
         
            \textbf{Documentation} & G.1  & \checkmark & \checkmark & \checkmark & \checkmark & \checkmark & \checkmark & \checkmark & \checkmark & \checkmark & \checkmark & \checkmark & 11 \\ \hline
            \textbf{Documentation} & G.2 & \checkmark & \checkmark & \checkmark & \checkmark & \checkmark & \checkmark & \checkmark & \checkmark & \checkmark & \checkmark & \checkmark & 11 \\ \hline
            \textbf{Documentation} & G.3  & \checkmark &  & \checkmark &  & \checkmark &  &  & \checkmark &  & \checkmark & \checkmark & 6 \\ \hline
            \textbf{Documentation} & G.4  & \checkmark &  & \checkmark & \checkmark &  &  &  & \checkmark & \checkmark & \checkmark & \checkmark & 7 \\ \hline
            \textbf{Documentation} & G.5  & \checkmark &  & \checkmark & \checkmark &  &  &  & \checkmark & \checkmark & \checkmark & \checkmark & 7 \\ \hline
            \textbf{Documentation} & G.6  & \checkmark &  & \checkmark &  &  &  &  &  & \checkmark & \checkmark & \checkmark & 5 \\ \hline
            \textbf{Documentation} & G.7  &  &  & \checkmark &  &  &  &  & \checkmark & \checkmark & \checkmark &  & 4 \\ \hline
            \textbf{Documentation} & G.8  &  &  & \checkmark &  &  &  &  &  & \checkmark & \checkmark & \checkmark & 4 \\ \hline
            \textbf{Documentation} & G.9  &  &  & \checkmark &  &  &  &  &  &  &  &  & 1 \\ \hline
            \textbf{Documentation} & G.10 & \checkmark &  & \checkmark &  & \checkmark &  &  & \checkmark &  & \checkmark & \checkmark & 6 \\ \hline
            \textbf{Documentation} & G.11 & \checkmark &  & \checkmark & \checkmark  &  &  &  &  &  & \checkmark & \checkmark & 5 \\ \hline
            \textbf{Documentation} & G.12 & \checkmark &  & \checkmark &  &  &  &  &  &  & \checkmark & \checkmark & 4 \\ \hline
            \textbf{Documentation} & G.13 & \checkmark & \checkmark &  & \checkmark & \checkmark &  &  &  &  &  &  & 4 \\ \hline

            \textbf{Planning} & G.14  & \checkmark &  & \checkmark & \checkmark &  &  &  & \checkmark & \checkmark & \checkmark & \checkmark & 7 \\ \hline
            \textbf{Planning} & G.15  & \checkmark & \checkmark & \checkmark & \checkmark &  &  &  & \checkmark & \checkmark & \checkmark & \checkmark & 8 \\ \hline
            \textbf{Planning} & G.16  & \checkmark &  & \checkmark & \checkmark &  &  &  &  & \checkmark & \checkmark & \checkmark & 6 \\ \hline
            \textbf{Planning} & G.17  & \checkmark &  & \checkmark & \checkmark &  &  &  &  &  & \checkmark & \checkmark & 5 \\ \hline
            \textbf{Planning} & G.18  & \checkmark &  & \checkmark & \checkmark &  &  &  & \checkmark & \checkmark & \checkmark & \checkmark & 7 \\ \hline
            \textbf{Planning} & G.19  & \checkmark &  & \checkmark & \checkmark &  &  &  & \checkmark &  & \checkmark & \checkmark & 6 \\ \hline
            \textbf{Planning} & G.20  & \checkmark &  & \checkmark & \checkmark &  &  &  & \checkmark & \checkmark & \checkmark & \checkmark & 7 \\ \hline
            \textbf{Planning} & G.21  & \checkmark &  & \checkmark & \checkmark &  &  &  & \checkmark & \checkmark & \checkmark & \checkmark & 7 \\ \hline

            \textbf{Operation} & G.22 &  &  & \checkmark & \checkmark &  &  &  & \checkmark &  &  &  & 3 \\ \hline
            \textbf{Operation} & G.23 & \checkmark & \checkmark & \checkmark & \checkmark &  & \checkmark &  & \checkmark &  & \checkmark & \checkmark & 8 \\ \hline
            \textbf{Operation} & G.24 & \checkmark & \checkmark & \checkmark &  &  & \checkmark & \checkmark & \checkmark & \checkmark & \checkmark & \checkmark & 9 \\ \hline

            \textbf{Analysis} & G.25 & \checkmark &  & \checkmark & \checkmark &  &  &  &  &  &  & \checkmark & 4 \\ \hline
            \textbf{Analysis} & G.26 & \checkmark &  & \checkmark & \checkmark &  &  &  & \checkmark &  &  & \checkmark & 5 \\ \hline
            \textbf{Analysis} & G.27 & \checkmark &  & \checkmark & \checkmark &  &  &  & \checkmark &  & \checkmark & \checkmark & 6 \\ \hline
            \textbf{Analysis} & G.28 & \checkmark &  & \checkmark & \checkmark &  &  &  &  &  & \checkmark & \checkmark & 5 \\ \hline
            \textbf{Analysis} & G.29 & \checkmark &  & \checkmark & \checkmark &  &  &  & \checkmark &  & \checkmark & \checkmark & 6 \\ \hline
            \textbf{Analysis} & G.30 & \checkmark &  & \checkmark &  &  &  &  & \checkmark &  &  & \checkmark & 4 \\ \hline
            \textbf{Analysis} & G.31 &  &  & \checkmark &  &  &  &  &  &  &  &  & 1 \\ \hline
            \textbf{Analysis} & G.32 & \checkmark &  & \checkmark &  &  &  &  &  &  &  & \checkmark & 3 \\ \hline
            \textbf{Analysis} & G.33 & \checkmark &  & \checkmark &  &  &  &  &  &  &  &  & 2 \\ \hline
            \textbf{Analysis} & G.34 & \checkmark &  & \checkmark &  &  &  &  &  &  &  & \checkmark & 3 \\ \hline
            
            \textbf{Documentation} & G.35 &  &  &  &  &  & \checkmark &  &  &  &  &  & 1 \\ \hline
            \textbf{Documentation} & G.36 & \checkmark &  &  &  &  & \checkmark & \checkmark &  &  &  &  & 3 \\ \hline
            \textbf{Documentation} & G.37 & \checkmark &  &  &  & \checkmark & \checkmark & \checkmark & \checkmark & \checkmark & \checkmark & \checkmark & 8 \\ \hline

            \multicolumn{2}{c|}{\textbf{Count}} & 31 & 6 & 33 & 21 & 6 & 7 & 5 & 21 & 15 & 25 & 29 & --- \\ \hline \hline       
        \end{tabular}
    }
\end{table*}

In Tables \ref{tab:relacao_rr_diretrizes} and \ref{tab:relacao_diretrizes_rr}, it is possible to observe that RR.3 maps a total of 33 experimental guidelines (33 out of 37), surpassing the 31 guidelines (31 out of 37) mapped in RR.1. This higher coverage reflects the complexity and diversity of aspects addressed in qualitative studies. The guidelines cover various phases of the experimental process, including documentation, planning, operation, and analysis, demonstrating the comprehensiveness of the items incorporated into RR.3.

Given this adherence, we will now delve deeper into RR.3 and its relationship to the guidelines. RR.3 is characterized in six main phases of the research process.

\textbf{Metadata:} includes the record title, a brief project description, a list of contributors, study category, licensing type, a list of relevant subjects, and associated tags.
  
\textbf{Study Information:} specifies the general objectives of the research and the questions guiding the study. Objectives may be broken down into more specific investigations, and the estimated project duration should be indicated, including expected start and end dates.
  
\textbf{Design Plan:} characterized succinctly, including a brief explanation of the meaning behind the study label. Additionally, sampling strategies and case selection should be described in detail, along with a justification for their choice.
  
\textbf{Data Collection:} identifies the sources and types of data to be used, distinguishing between pre-existing data and original data to be collected or generated during the research. Data collection methods should be thoroughly described, along with the tools and instruments to be used. Criteria for stopping data collection should also be clearly explained, taking into account factors such as data saturation or resource constraints.
  
\textbf{Analysis Plan:} specifies the approach and details of the data analysis process, including justifications for the choice of this approach in relation to the study objectives. The analysis process should include information on who will be involved, the procedures to be used, and any tools or software to be employed.
  
\textbf{Miscellaneous:} includes an optional reflection on the researcher's position relative to the phenomenon under study. This may encompass academic and personal positions, assumptions, values, and potential conflicts of interest.

In RR.3, the process begins with filling out the ``Metadata'' details, using guideline G.1 for the study title. Subsequently, the author's contact information is included in accordance with G.2, and relevant keywords are presented in accordance with G.9. The study abstract highlights the importance of the research, aligning with G.3, and the questions the research aims to address are defined in line with G.4.

In the ``Study Information'' section, the research problem is outlined in accordance with G.10, along with the research objectives, which should be formulated in accordance with G.11. The research context and factors influencing the generalizability of the results are considered, in accordance with G.12.

The ``Design Plan'' includes the formalization of research objectives and the definition of essential constructs according to G.14. The definition of the population from which the sample will be drawn is addressed per G.15. The study design, including the study type and sampling strategies, is described, aligning with G.19 and G.3.

In the ``Data Collection'' section, data sources and types are identified, following G.16. Data collection methods are described according to G.12. The conduct of data collection is detailed in alignment with G.20. Any deviations from the original plan are recorded per G.24. Additionally, the tasks that experiment subjects must perform are outlined, in accordance with G.17.

The ``Analysis Plan'' covers the description of the data analysis procedures following G.21. This section includes descriptions of statistical models and inference criteria, aligning with G.17 and G.10, and discusses the results of descriptive statistics to be reported, in line with G.25. Additionally, criteria for excluding data points or samples and plans for exploratory data analysis are considered, following G.22 and G.27.

The interpretation of results includes an explanation in relation to previous research, per G.28, and a discussion on the validity and potential biases of the results, aligning with G.29. General conclusions based on the obtained results are presented in line with G.32, and suggestions for future work are offered per G.34.

\subsection{Lessons Learned and Limitations}
\label{sec:lessons}

The concept of removing obstacles to the reproducibility and transparency of research findings and artifacts is particularly challenging in the field of computer science, especially in software engineering. Therefore, we recognize that investigating RRs is a way to make these practices achievable.

Despite RR being a concept still underutilized in software engineering, we have learned that it ensures that research data are not the sole focus of a study. Instead, it emphasizes the study's design, particularly before the survey is carried out. This change of focus helps to mitigate several common threats to validity in empirical studies, such as selective reporting, data dredging, and post hoc hypothesis formulation. In other words, the process promotes a culture of planning and transparency rather than relying solely on the persuasiveness of results.

Another important lesson is that Registered Reports foster accountability and dialogue with the research community from the earliest stages of the research process. When effectively implemented, phase 1 review encourages authors to justify methodological choices, define measurable variables, and provide a clear rationale for hypotheses in advance. This also creates opportunities for reviewers to provide constructive feedback before resources are spent, ultimately improving study quality.

Nevertheless, we also observed that Registered Reports do not guarantee reproducibility or replicability. They are best suited for hypothesis-driven studies, where experimental protocols and statistical analysis plans can be pre-registered. In exploratory or design-oriented studies, which are common in software engineering, the RR format can be less directly applicable. In such cases, adaptations or complementary practices such as structured metadata and artifact sharing are necessary.

The main limitations can be summarized as follows:

\textbf{Lack of protocol transparency:} It is essential to ensure that readers and reviewers can verify whether the approved protocol was strictly followed. However, some major journals do not require RRs for the initial phase of a study due to time constraints. This creates a risk of partial adherence to protocols and reduces trust in the RR process, particularly in short-term or fast-paced research projects.

\textbf{Lack of standardization:} RRs are often written inconsistently, with missing details or without explicit hypotheses. There is still no widely adopted standard template for RR writing in software engineering, which leads to heterogeneity and difficulties in automating or streamlining the process. A metadata-driven approach and integration with open science tools could help to standardize RR content and ensure completeness.

\textbf{Delay and bureaucracy in phase 1:} The review of protocols in phase 1 may take several months, delaying the start of the actual study. For researchers constrained by funding cycles, academic deadlines, or short project durations, this is a significant obstacle. In addition, the need for ethics committee approval, particularly in studies involving human participants, can further extend the timeline, sometimes making RRs infeasible.

\textbf{Limited coverage of research types:} RRs are well aligned with confirmatory, quantitative studies, but they are less naturally suited to qualitative, exploratory, or mixed-methods approaches. This restricts their applicability in many areas of software engineering, which often rely on case studies, grounded theory, or design science research.

In relation to OSF, it provides different RR types, which present an opportunity for detailed and transparent study conception documentation, as it plans to conduct final results checking. We learned that the RR types we chose for this study are quite adequate for data-driven studies, where data are collected through measurable variables and testable hypotheses are formulated using statistical techniques. Therefore, for the guidelines by \citet{Jedlitschka_et_al2008}, we observed that most of them are considered in the analyzed RR types. However, we understand that combining RR.1 and RR.3 would be necessary to address the experiment guidelines fully. Unfortunately, OSF does not allow this customization.

Another limitation is the absence of a dedicated RR type on the OSF platform specifically focused on Software Engineering, particularly considering its research artifact types, as discussed by \citet{OliveiraJr_et_al2024}. For instance, code repositories, UML diagrams, models, or scripts, which are central to reproducibility in SE, are not explicitly supported in the current RR templates.

Furthermore, none of the existing OSF RR types fully meet all the guidelines outlined by \citet{Jedlitschka_et_al2008}. As a result, we recommend proposing a customization or the development of new RR types in collaboration with OSF. This would enable closer alignment with SE-specific experimentation practices, improve artifact documentation, and reduce gaps between methodological planning and artifact reproducibility.

In summary, we learned that RRs contribute positively to research integrity and planning but face significant challenges regarding adoption, standardization, and adaptation to the specific needs of software engineering. Their future success in this field will depend on broader community engagement, support for tools, and closer alignment with the diverse methodological approaches used in SE research.

Overall, the lessons learned show that RRs enhance transparency, accountability, and methodological rigor. At the same time, the limitations highlight the practical, organizational, and disciplinary barriers that must be addressed for their effective implementation. This dual perspective highlights the importance of integrating methodological innovation with infrastructural and cultural changes to make RRs fully effective in software engineering.


\section{Prospective Actions}
\label{sec:actions}

We provide a set of future actions to be taken in light of this work, with the overarching goal of enhancing the maturity, adoption, and usability of RRs for controlled experimentation in Software Engineering. These actions are not isolated; instead, they form a roadmap that bridges methodological rigor, community acceptance, and tool-supported reproducibility.

The first step is to \textbf{review the RR types} based on the guidelines by \citet{Jedlitschka_et_al2008} and incorporate items to fill relevant missing elements. This will enhance and improve RR experiment documentation by ensuring that crucial aspects, such as context, threats to validity, and reproducibility requirements, are not overlooked. The review process will also help us distinguish between mandatory and optional elements, clarifying which parts of the RR structure should be emphasized for controlled experimentation in Software Engineering.

Another critical action is to \textbf{identify which RR elements are required, or at least suggested, by journals and conferences} that accept RR submissions. This mapping exercise will serve as a bridge between our proposed structure and the actual publication requirements. It will also serve as a valuable resource for the community, offering examples of how each RR item is written in practice. By analyzing differences across venues, we can better guide researchers on tailoring their submissions without sacrificing methodological transparency.

We will also \textbf{inquire with OSF about the option to customize the existing Registered Reports or to create a new one}. Having a dedicated template or customized workflow within OSF could significantly speed up the process of writing controlled experiment RRs. If such customization is feasible, we will seek to incorporate RR elements that span a variety of software engineering artifacts, including code repositories, diagrams, scripts, and domain-specific languages. Additionally, we will provide explicit guidance on how to reuse or reproduce these artifacts, thereby strengthening the reproducibility chain across experiments.

We foresee the need to \textbf{identify existing RR standards or patterns} for controlled experimentation in established fields such as Medicine and Psychology. These disciplines have a longer tradition of registered studies and could provide ready-made structures or best practices that can be adapted to the context of Software Engineering. By leveraging domain expertise from these fields, we aim to avoid reinventing the wheel and instead build upon proven strategies to formalize and accelerate RR writing in our community.

The \textbf{creation of RR metadata for interchange} among tools and environments is another prospective action. At present, there is no unified, automated solution for representing and exchanging RR data. Developing a standard metadata schema would facilitate the integration of RRs across different platforms, including repositories, submission systems, and experiment management tools. Such metadata would also support automated validation, indexing, and long-term preservation, in line with open science principles.

To standardize the RR description language and structure for experiments, we will \textbf{examine the compatibility of existing software engineering-related RRs}. This comparative analysis will highlight overlaps, gaps, and inconsistencies, helping us refine our own template. Moreover, \textbf{analyzing existing Data Management Plans (DMPs)} will provide insights into how data generated or reused in experiments should be curated, preserved, and documented. Since DMPs focus on data life cycle management and provenance, they are highly complementary to RRs. In addition, considering categories and elements of research software \citep{Felderer_et_al2025} may further enrich RR descriptions by clarifying how software artifacts are documented, cited, and reused.

Another important line of action is to \textbf{analyze the human-related aspects required by ethical committees}. Controlled experiments in Software Engineering often involve participants such as students, professionals, or volunteers. By explicitly considering ethical requirements, particularly when sensitive data are collected or when procedures may include embarrassment, stress, or other potentially traumatic experiences, we can improve the robustness of RR descriptions and ensure compliance with institutional review boards and ethical guidelines.

Finally, as an essential task, we \textbf{plan to conduct an empirical study} focused on writing RRs by individuals engaged in Software Engineering research, such as students and instructors in Empirical Software Engineering courses. Using our proposed template, participants will produce RRs that will then be submitted for peer review. We will analyze reviewer feedback to identify strengths and weaknesses of the template. As part of the study, participants may also be asked to write RRs without the template, enabling a comparative evaluation across dimensions such as completeness, clarity, and methodological rigor. This empirical validation will not only refine our approach but also provide evidence-based recommendations for adopting RRs within the Software Engineering community.

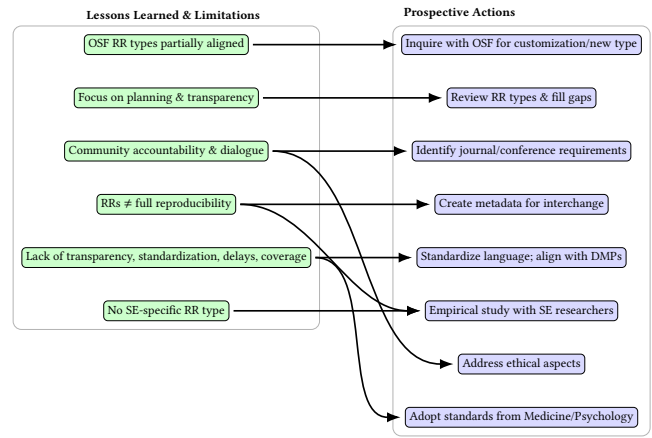
\begin{figure}[!h]
\centering
\resizebox{.48\textwidth}{!}{%
\begin{tikzpicture}[
  font=\scriptsize,
  >=Latex,
  every node/.style={align=left},
  lesson/.style={rounded corners, draw=black, fill=green!20, rounded corners=2pt, inner sep=2pt},
  action/.style={rounded corners, draw=black, fill=blue!15, rounded corners=2pt, inner sep=2pt},
  bridge/.style={->, thick, black, shorten <=2pt, shorten >=2pt},
  group/.style={draw=black!30, rounded corners=4pt, inner sep=4pt}
]

\node[lesson] (LL5) at (0, 2.7) {OSF RR types partially aligned};
\node[lesson] (LL1) at (0, 1.8) {Focus on planning \& transparency};
\node[lesson] (LL2) at (0, 0.9) {Community accountability \& dialogue};
\node[lesson] (LL3) at (0, 0.0) {RRs $\neq$ full reproducibility};
\node[lesson] (LL4) at (0,-0.9) {Lack of transparency, standardization, delays, coverage};
\node[lesson] (LL6) at (0,-1.8) {No SE-specific RR type};

\node[anchor=south west] at ([xshift=-2pt,yshift=4pt]LL5.north west) {\textbf{Lessons Learned \& Limitations}};

\node[action] (AC3) at (6.0, 2.7) {Inquire with OSF for customization/new type};
\node[action] (AC1) at (6.0, 1.8) {Review RR types \& fill gaps};
\node[action] (AC2) at (6.0, 0.9) {Identify journal/conference requirements};
\node[action] (AC5) at (6.0, 0.0) {Create metadata for interchange};
\node[action] (AC6) at (6.0,-0.9) {Standardize language; align with DMPs};
\node[action] (AC8) at (6.0,-1.8) {Empirical study with SE researchers};
\node[action] (AC7) at (6.0,-2.7) {Address ethical aspects};
\node[action] (AC4) at (6.0,-3.6) {Adopt standards from Medicine/Psychology};

\node[anchor=south west] at ([xshift=-2pt,yshift=4pt]AC3.north west) {\textbf{Prospective Actions}};

\node[group, fit=(LL5)(LL1)(LL2)(LL3)(LL4)(LL6)] {};
\node[group, fit=(AC3)(AC1)(AC2)(AC5)(AC6)(AC8)(AC7)(AC4)] {};

\draw[bridge] (LL1.east) -- (AC1.west);
\draw[bridge] (LL2.east) -- (AC2.west);
\draw[bridge] (LL3.east) -- (AC5.west);
\draw[bridge] (LL4.east) -- (AC6.west);
\draw[bridge] (LL5.east) -- (AC3.west);
\draw[bridge] (LL6.east) -- (AC8.west);

\draw[bridge] (LL2.east) to[out=0, in=180, looseness=0.9] (AC7.west);
\draw[bridge] (LL3.east) to[out=0, in=180, looseness=0.9] (AC8.west);
\draw[bridge] (LL4.east) to[out=0, in=180, looseness=0.9] (AC4.west);

\end{tikzpicture}
}
\caption{Lessons learned and limitations driving prospective actions.}
\label{fig:conceptual-model-lessons-actions}
\end{figure}

Figure~\ref{fig:conceptual-model-lessons-actions} shows a mind map connecting the lessons learned and limitations of our work on RRs with the prospective actions proposed for their improvement. The left side highlights key challenges, such as the need for transparency, accountability, and SE-specific standards. In contrast, the right side presents actions including refining RR types, aligning with publication requirements, OSF customization, metadata development, and conducting empirical studies. The directed edges highlight how each lesson informs a specific action, providing a roadmap that addresses current barriers while promoting reproducibility and rigor in Software Engineering experiments.

\section{Final Remarks}
\label{sec:conclusion}

This paper presented a proposal for software engineering controlled experiment registered reports, based on a set of guidelines by \citet{Jedlitschka_et_al2008}.

We compared all the OSF-registered report types and narrowed them down to four for further analysis. We selected one report type that aligns best with the experiment guidelines. We discussed how the chosen RR type maps to the guidelines. Additionally, we reveal the current limitations in customizing, adding, or creating new RR types under the OSF platform.

We have identified specific actions to write effective RRs for controlled experiments in software engineering. We recognize that establishing guidelines for RRs in software engineering is challenging yet essential, particularly considering the nature of the artifacts in this field.

\section*{Acknowledgments}

Edson OliveiraJr thanks CNPq/Brazil grant \#311503/2022-5.

\bibliographystyle{ACM-Reference-Format}
\bibliography{references}

\end{document}